\documentclass[twocolumn]{aastex63}

\shorttitle{Multimessenger Observations of T\.ZOs}
\shortauthors{DeMarchi et al.}

\graphicspath{{./}{figures/}}

\begin{document}

\title{Prospects for Multimessenger Observations of Thorne-\.Zytkow Objects}

\author[0000-0003-4587-2366]{Lindsay ~DeMarchi}
\affiliation{Center for Interdisciplinary Exploration and Research in Astrophysics and Department of Physics and Astronomy, Northwestern University, 2145 Sheridan Road, Evanston, IL 60208-3112, USA}
\author[0000-0001-7764-8030]{J.R. ~Sanders}
\affiliation{Marquette University,
Physics Department, 1420 W. Clybourn St.
Milwaukee, WI 53233, USA}
\author[0000-0003-2184-1581]{Emily M. Levesque}
\affiliation{University of Washington Astronomy Department,
Physics and Astronomy Building, 3910 15th Ave NE,
Seattle, WA 98105, USA}

\begin{abstract}
Thorne-\.Zytkow objects (T\.ZOs) are a class of stellar object comprised of a neutron star core surrounded by a large and diffuse envelope. Their exterior appearance is identical to red supergiants; the distinctive electromagnetic signature of a T\.ZO is a suite of unusual chemical abundance patterns, including excesses of Li, Rb, Mo, and Ca. However, electromagnetic observations cannot unambiguously identify the presence of a neutron star core. Detection of continuous gravitational wave emission from a rotating neutron star core would provide strong supporting evidence for the existence of T\.ZOs. We present a model for gravitational wave detector confirmation of T\.ZOs and demonstrate that these objects should be detectable with Advanced LIGO. We also investigate possible targets for joint optical and gravitational searches, and comment on prospects for detectability in both current and future gravitational wave detector networks. \\
\end{abstract}

\section{Introduction}
\subsection{What is a T\.ZO?}
Thorne-\.Zytkow objects (T\.ZOs) are a class of star originally proposed by \citet{ThorneZytkow1975,ThorneZytkow1977}, comprised of a neutron star (NS) core surrounded by a large and diffuse envelope. T\.ZOs are expected to form as a result of the evolution of two massive ($\gtrsim$8$M_{\odot}$) stars in a close binary, with the neutron star forming when the more massive member of the binary explodes as a core-collapse supernova. During the subsequent evolution of the system, the expanding envelope of the remaining star may lead to a common envelope state and the spiral-in of the neutron star into the core of its companion \citep{Taam1978}. Alternately, \citet{Leonard1994} propose that a T\.ZO may be produced when a newly-formed neutron star receives a supernova ``kick" velocity in the direction of its companion and becomes embedded. A T\.ZO could also form dynamically, with a newly-formed neutron star captured as a companion by a main-sequence star in a globular cluster \citep{Ray1987}.

T\.ZOs represent a completely new class of stellar object, offering a novel model for stable stellar interiors and a new evolutionary stage for binary massive stars. However, for nearly 40 years these objects existed solely as a theoretical class; there had never been a positive observational identification of a T\.ZO.  This is partly due to the challenges of detection: T\.ZOs were predicted to have an outward appearance virtually indistinguishable from that of red supergiants (RSGs). The only proposed observational signature of a T\.ZO was based on rare nucleosynthesis processes made possible by their interior structure, combining their completely convective envelopes, high convection speeds, and extreme temperatures and densities at the base of the envelope near the neutron star. This in turn was predicted to give rise to an unusual chemical abundance pattern in the star's atmosphere showing excesses of elements such as Li, Rb, Mo, and Ca (e.g. \citealt{Cannon1993,Biehle1994,Podsiadlowski1995,Tout2014}).

\citet{Levesque2014} identified the first observational evidence for the confirmed existence of T\.ZOs in the form of HV 2112, a supergiant in the Small Magellanic Cloud \citep{Worley2016,McMillanChurch2018} whose atmospheric chemistry displayed the precise elemental signature predicted by theoretical models of T\.ZO interiors with enhanced quantities of Li, Rb, Mo, and Ca evident in high-resolution spectroscopy. The star was previously classified as a red supergiant but displayed a particularly cold temperature and high luminosity, also in agreement with the predictions of T\.ZO models. Finally, it showed evidence of unusual hydrogen emission in its spectrum, consistent with the properties of an atmosphere excited by shocks generated by stellar pulsations.

HV 2112 is by far the most compelling candidate ever observed for confirmation of the T\.ZO model; however, it is not impossible that this star could simply be a (very extreme) chemical anomaly or some other previously-unexplored class of massive or highly luminous star (e.g. \citealt{OGrady2020}). Open questions such as the rate of neutrino cooling in accreting neutron stars and its effect on the stability and final fate of the T\.ZO model (e.g. \citealt{Fryer96,Yakovlev03}) also make understanding the rarity and lifetimes of T\.ZOs (and, as a result, available observable populations) unclear. At a fundamental level the T\.ZO model can only be proved through the direct observational confirmation of the presence of a neutron star core inside a red supergiant. While this is effectively impossible for electromagnetic observations (as any signature of the neutron star is expected to be completely obscured by the surrounding envelope), the burgeoning field of gravitational wave astronomy offers an exciting new window into these strange stars.

\subsection{Why Are They Multi-Messenger Candidates?}
T\.ZOs are excellent targets for multi-messenger astronomy. The neutron star asymmetries and rapid rotation speeds required for producing continuous or periodically continuous gravitational waves are expected as a consequence of the central neutron star's evolution in the T\.ZO core - it will start as an extremely rapid rotator with accretion-induced asymmetries and spin down dramatically over the course of the T\.ZO lifetime (e.g. \citealt{Podsiadlowski1995}). \cite{Liu2015} proposed that the X-ray source 1E161348-5055, with a 6.67-hour period, is the descendant of a T\.ZO; in their scenario the neutron star's rotation rate is dramatically slowed by a coupling between its magnetic field and the former RSG envelope, with the envelope then ejected by stellar winds or sporadic pulsational mass loss. Both \cite{Liu2015} and \citet{Podsiadlowski1995} posit the spin-down of a T\.ZO's central neutron star due to interactions in the stellar interior, as well as inherent asymmetries in the slowly-accreting neutron star itself.

Past work explored the possibility of a transient gravitational wave signal produced during the formation of T\.ZOs from the merger of a neutron star and the dense stellar core of the RSG. \citet{Nazin1995} modeled a number of different merger scenarios and predicted maximum strains of $h\sim10^{-24}$ at frequencies ranging from $10^{-5}$ to 0.1 Hz. Unfortunately, these frequencies are too low for detection with aLIGO and the strain is too weak for detection with LISA. This method of multi-messenger T\.ZO detection would also be exceptionally challenging, as it would entail capturing the brief transient gravitational wave signal from the moment of formation and identifying a newly-formed T\.ZO spectroscopically, an unlikely prospect given that surface abundance changes will likely be undetectable immediately after T\.ZO formation (e.g. \citealt{Cannon1993}).

Unlike an electromagnetic signal, the propagation of a continuous gravitational wave signal from a T\.ZO's central neutron star through spacetime will be unhindered by the surrounding stellar envelope. As a result, a true post-formation T\.ZO should have an unmistakable and persistent multi-messenger signature - the electromagnetic properties of its red supergiant-like envelope combined with the gravitational wave emission of its central neutron star. With existing data from aLIGO - and the potential for more sensitive observations from future gravitational wave observatories such as the space-based Evolved Laser Interferometer Space Antenna (eLISA) - this offers the possibility of a simple and efficient way to confirm the status of existing T\.ZO candidates and carry out large-scale searches for future populations of T\.ZOs.

Estimating the current detectable population in the Milky Way requires considering both the formation rates and lifetimes of T\.ZOs. Using the \citet{Taam1978} and \citet{Leonard1994} formation scenarios for T\.ZOs, \citet{vanParadijs1995} predicted that these objects should form at a rate of $\sim10^{-4}$ yr$^{-1}$ in the Milky Way; \citet{Podsiadlowski1995} estimated a slightly higher rate of $\sim 2 \times 10^{-4}$ yr$^{-1}$. More recently, \citet{Tout2014} calculated a rate based on binary stellar evolution models, finding that among massive star binaries about 2\% of these systems should form T\.ZOs. Similarly, \citet{Neugent2020} estimated that 2.42 $\pm$ 0.01\% of RSGs in the Large Magellanic Cloud have compact companions, wherein the first high-mass X-ray binary comprised of a RSG and a neutron star was recently discovered \citep{Hinkle2020}. Collectively, while any estimates of T\.ZO formation rates will be complicated by uncertainties surrounding the binary and merger fraction of massive stars as a function of metallicity and evolutionary state, these four estimates are all broadly consistent. 

If T\.ZO lifetimes are primarily dictated by mass loss (a conclusion supported by their presumed luminous cool states and the observed properties of HV 2112), \citet{vanParadijs1995} concluded that T\.ZO should live for $\sim10^5$-$10^6$ years, suggesting a population of 10-100 observable T\.ZOs in the Milky Way; \citet{Podsiadlowski1995} estimates 20-200. \citet{Tout2014} estimate a slightly shorter lifetime of $10^4$ years but note that uncertainties in the rate of T\.ZO formation and lifetime also depend strongly on the envelope mass lost during the merging and spiraling-in process (see also \citealt{Liu2015}). 

Overall, estimates suggest as many as 100 T\.ZOs in the Milky Way. By calculating their expected persistent gravitational wave signatures, carrying out targeted gravitational wave searches in regions with large populations of RSGs (such as clusters rich in evolved massive stars), and combining any detections with electromagnetic follow-up observations, we should be capable of definitively identifying T\.ZOs as multi-messenger sources.

\section{Observable Signatures}
\subsection{Electromagnetic Signature}

While a T\.ZO might masquerade as a red supergiant, the neutron star at its core would cause a unique spectroscopic signature at the star's surface impossible without the unusual combination of a neutron degenerate core and a large convective envelope with high convection speeds. The hallmarks of this process would include a suite of metal abundances produced by the interrupted rapid proton (irp-)process that is unique to T\.ZOs, such as Rb I and Mo I (e.g. \citealt{Cannon1993, Biehle1994,Levesque2014}), as well as Li$^7$ \citep{Podsiadlowski1995} and Ca I \citep{Tout2014}.

It is worth noting that some of these individual abundance enhancements can also be produced by the $s$-process; for example, Rb and Mo have been separately observed in $s-$process-enhanced stars, and enhanced Li has been observed in some AGB stars. Increased abundances in any one element are not sufficient to identify a cool luminous star as a candidate T\.ZO. The unique combination of all of these element enhancements (and a lack of other $s$-process abundance signatures, such as enhanced Ba) are required, such as in the optical spectrum of HV 2112, a strong argument in favor of its status as a true T\.ZO candidate. However, it is also possible that these enhancements could be explained as a super-asymptotic giant branch (SAGB) star combining several never-before-seen abundance effects, such as $s$-process enhancement of Rb, Li production at the base of an SAGB star convective envelope, and some novel explanation for HV 2112's Ca excess (e.g. \citealt{GarciaHernandez2013,Tout2014}). \citet{OGrady2020} similarly argue that HV 2112 could potentially be an SAGB star based on comparisons of its lightcurve with other cool luminous stars.

Beyond the spectroscopic signature of a T\.ZO, extremely powerful winds are anticipated. Strong dust-driven winds and episodic mass loss are both common in cool luminous stars, but T\.ZO winds should be further enhanced by the coupling of the neutron star's magnetic field to its envelope as well as the fact the Alfven radius (the co-rotation radius of material) is larger than that of the compact core. This can disrupt the envelope with powerful bursts as the core spins down. Due to this magnetic braking, a T\.ZO late in its life would look like a compact object with a very thin SN-remnant-like shell, a result of the outer layers of the star being stripped away by powerful winds. The compact object will have also slowed down to an anomalously long spin period. Additionally, the object would have an unusual proper motion relative to its surroundings, a consequence of the compact object's merger with the massive companion (whether this proper motion is slower or faster depends on the orbital dynamics of the original binary system). This is the model invoked by \citet{Liu2015} to explain the X-ray source 1E161348-5055 as a slowly-rotating neutron star produced as the end result of a T\.ZO.

However, identifying {\it post}-T\.ZO neutron stars is of only limited value when searching for the T\.ZOs themselves, and both episodic mass loss and atypical proper motions in cool luminous stars can be explained by other phenomena. Combined, while T\.ZOs are expected to display a unique suite of electromagnetic signatures, truly convincing evidence for the presence of a neutron star core requires some unambiguous signature of the neutron star itself.

\subsection{Gravitational Wave Signature}
Gravitational waves are produced by any object showing time-varying accelerations of quadrupolar mass distributions. Because larger masses create larger perturbations to spacetime, it is no surprise the first class of observed gravitational waves were compact binary inspirals, as they are also well-modeled and comparatively high-strain gravitational wave signals. Thorne-\.Zytkow objects can be considered a failure mode of the formation of a compact binary object, with mass dynamics and densities significant for consideration in gravitational wave generation. The gravitational wave signal of T\.ZO formation was investigated by \cite{Nazin1995}. In the detector basis, this would present as an unmodeled short gravitational wave signal, or burst, at frequencies ranging from $10^{-5}$ to $\sim 0.1$ Hz with characteristic strain $h \sim 10^{-23.5}$ (for a T\.ZO forming 10 kpc from the Sun). These frequencies are inaccessible to terrestrial detectors due to the impact of seismic noise, and are also too low in amplitude to appear in the next generation of space-based detectors.

\textit{After} a Thorne-\.Zytkow object has formed, the gravitational-wave emission will be dominated by the spin evolution of its central neutron star. The central neutron star will have residual rotation due to conservation of angular momentum during its formation, and the braking of the neutron star will result in some energy loss into gravitational waves. The maximum possible gravitational wave luminosity, or spindown limit, is set by the energy lost in the slowing of this rotation. From a gravitational wave perspective, a post-formation Thorne-\.Zytkow object is indistinguishable from a rotating isolated neutron star. The methodologies developed for searching for gravitational waves from supernova remnants with no associated electromagnetic pulsations \citep{CasAMethods,Prix2009} can be used to investigate if such an isolated neutron star is co-located with an apparent red supergiant.

In the event that an isolated neutron star appears at an appropriate range of frequencies for terrestrial detectors, but at a strain too weak to rise above the noise, a non-detection would set limits on gravitational wave emission under the chosen search model.

\section{Where to Look}\label{EMSearch}
As T\.ZOs are likely to form from RSG+NS binaries and to be mistaken for RSGs in electromagnetic surveys, the best place to search for T\.ZOs is in RSG-rich regions or clusters. \citet{Levesque2014} focused on M-type (i.e. cold) RSGs with well-established physical properties in the Milky Way and Magellanic Clouds for their high-resolution spectroscopic search. However, in a gravitational wave search for T\.ZOs - more limited in distance but unencumbered by problems such as interstellar dust - the most compelling targets would be the RSG-rich stellar clusters of the inner Milky Way.

There are six RSG-rich clusters near the base of the Scutum-Crux arm: 
RSGC1 ($d=6.6\pm0.89$ kpc, \citealt{Davies2008}); 
RSGC2 ($d=5.83^{+1.91}_{-0.76}$ kpc, \citealt{Davies2007}); 
RSGC3 ($d=5.9\pm0.3$ kpc, \citealt{GFN2012}); 
Alicante 10 (a distinct cluster of RSGs associated with RSGC3 but with a closer center of $d=5.1\pm0.2$ kpc, \citealt{GFN2012});
RSGC4 ($d \sim 6.6$ kpc, \citealt{Negueruela2010}); 
and RSGC5 ($d \sim 6$ kpc, \citealt{Negueruela2011}). 
While the extinction in the direction of these clusters is high and makes optical spectroscopy challenging, a search for the gravitational wave signatures of T\.ZOs targeting these clusters would be an ideal way to focus the search on a large population of RSGs.

RSGC1 is at a distance of 6.6$\pm$0.89 kpc and fairly compact with a radius of r$\sim$1.5' \citep{FroebrichSholz2013}. According to \citet{FroebrichSholz2013} the cluster is 10Myr old, with 210 massive stars  and 14 confirmed RSG members \citep{Davies2008}. RSGC1 also counts one X-ray bright pulsar as a member, associated with a TeV gamma-ray source and described as a rotation-powered pulsar with an age of $\sim$23 kyr \citep{GotthelfHalpern2008}. This, combined with the 10Myr cluster age, is a powerful suggestion that at least some members of the cluster have already undergone core collapse and produced neutron stars that could be either binary companions or merged cores for the remaining RSGs. The cluster is also associated with a number of other X-ray and high-energy sources \citep{Figer2006,Townsley2018}; though none have been directly spatially associated with a RSG, \citet{Figer2006} interpret them as evidence of supernova activity in the region.

RSGC2 and RSGC3 are both similarly spatially compact, with $r\sim$1.8' \citep{FroebrichSholz2013}. Both are estimated to have 115 massive stars; RSGC2 has 26 RSGs while RSC3 has 16, and the clusters are estimated at 17Myr and 20Myr old, respectively. While there are fewer stars in both these clusters, the older ages combined with the presence of remaining RSGs is a strong suggestion that there should be neutron stars present in the clusters, the product of past core-collapse activity.

The remaining three RSG clusters described above - Alicante 10, RSGC4, and RSGC4 - are all considerably more diffuse ($\sim$6-7$'$), and each contains roughly a dozen observed/candidate RSG members. 

From this group, RSGC1 stands out as the most compelling first target for a gravitational-wave-based search for T\.ZOs due to its small angular size, its large massive star and RSG population, the confirmed presence of at least one neutron star member, and its estimated age: at 10Myr it is old enough to host neutron star members but young enough that we could potentially detect a T\.ZO earlier in its spin-down process.

\section{GW Calculation}
Searches for gravitational waves from hypothetical neutron stars with known sky position and unknown ephemeris have significant precedent, with applications to neutron-star rich sky locations such as the galactic center \citep{GalacticCenter} and supernova remnants such as Cassiopeia A \citep{CasAMethods}. Gravitational wave interferometers have an antenna pattern that covers nearly all of the sky simultaneously \citep{Schutz2011} with no need for instrumental pointing during data taking. Instead, archival data is demodulated to remove the Doppler shift of the Earth's motion relative to a sky position of interest, retroactively ``pointing" the data \citep{Patel2010}. 

The model for gravitational wave emission from rotating neutron stars assumes a time-varying spin of an object with a given ellipticity $\epsilon$. This results in a model of the gravitational wave strain, $h(f)$, as a slowly varying sinusoid of a frequency integer multiple of the rotation frequency. The primary mode of emission is the $l = m = 2$ mass quadrupole mode such that $f_{\rm GW} = 2f_{\rm{rotation}}$ \citep{PhysRevD.20.351}. The impact of energy loss on the neutron star rotation rate is approximated through Taylor expansion with key variables of frequency $f$, spindown $\dot{f}$, and first spindown derivative $\ddot{f}$.

The canonical significance threshold for gravitational wave searches is the \emph{spindown limit}, the strain amplitude assuming that all observed change in rotation in the neutron star system is due to energy loss from gravitational wave emission, 

\begin{equation}\label{hspindown}
h_{\rm{spindown}} = \frac{1}{D} \sqrt{\frac{5GI_{zz}(-\dot{f})}{2c^3f}}
\end{equation}

where $h_{\rm{spindown}}$ is the gravitational wave strain tensor amplitude, $D$ is the distance to the source, $G$ is the gravitational constant, and $c$ is the speed of light. By its nature, this is an absolute upper limit; energy losses due to diverse non-gravitational wave mechanisms are readily observed in neutron star systems. As an example, current upper limits from searches for the Crab Pulsar in Advanced LIGO data place this gravitational wave energy fraction at less than 0.01\% of the total energy loss \citep{CrabVela2020}.

A neutron star with observed electromagnetic pulsations would have known values for $f$ and $\dot{f}$. In absence of a known emphemeris, limits must be set across a broad range of potential $f/\dot{f}$ pairs; effectively, stating limits on gravitational emissions at each potential frequency. Conventionally, the rotational frequency and spindown are assumed to have a power law relationship, $\dot{f} \propto f^n$, defining the \emph{braking index} $n$. The theoretical braking index for different emission models ranges from $n=2$ (pure dipole radiation) to $n=5$ (gravitational emission from a mass quadrupole) to $n=7$  (exotic treatments of magnetic fields). Observed ephemerides range from $n=2$ to $n=3$, indicating realistic pulsar emission universally dominated by electromagnetic radiation. As the gravitational wave signal is determined by the observed rotational frequency, not the idealized gravitar rotational ephemeris, a characteristic spindown of $n=2$ is an appropriate upper limit for realistic systems. Throughout this discussion, we focus on the $n=2$ braking index, as this is the most conservative of the upper limits and in good agreement with prior work on the braking index and gravitational wave signals from isolated neutron stars \citep{Abadie_2010}.

If we assume that this rate of spin evolution remains constant, we can define the characteristic age $\tau$ of the neutron star with a first-order Taylor expansion,

\begin{equation}\label{tau}
 \tau \approx
 \frac{1}{n-1} \left( \frac{f}{-\dot{f}}\right)
\end{equation}

The characteristic age is an age-like quantity which correlates well with real age for young neutron stars and is used for setting search limits on $\dot{f}$. A typical lower limit for characteristic age is 300 years, under the assumption that neutron-star-producing events in the galaxy within the past 300 years would have been associated with a recorded supernova.

The multi-messenger observations of gravitational waves from binary neutron star merger GW170817 have resulted in some additional constraints on potential neutron star equations of state. 

For the purposes of this analysis, we considered a neutron star with a canonical mass of $1.4 M_{\odot}$, assumed that all rotational inertia was along the $I_{zz}$ axis, and reviewed the equation of state models not ruled out by GW170817 \citep{Haster:2020sdh}. Stiffer equations of state allow for higher moments of inertia at a given mass, so for the purposes of upper limit calculations, the hardest equation of state not ruled out by GW170817, H4 \citep{Zhou:2017hfm}, was used for setting the numerical upper limit. As a reference, the softest equation of state, WFF1 \citep{Breu:2016ufb}, reduces the strain by about 20\%.

At current instrumental performance, continuous gravitational waves from a T\.ZO candidate like HV 2112 in the SMC ($\sim$60kpc) are not distinguishable from instrumental noise. Therefore, the proposal of searching Galactic red supergiants for potential new candidate T\.ZOs is driven by the limits of gravitational wave interferometer performance. Fortunately, there are several interesting supergiant clusters within the design sensitivity of the currently operating Advanced LIGO network. Of the six considered in Section \ref{EMSearch}, we highlight in particular RSGC1. The reported distance of $6.6\pm0.89$ kpc from \citet{Davies2008} is accessible to the current generation of detectors, and the confirmed observation of 14 RSGs and at least one neutron star shows evidence of all ingredients necessary for T\.ZO creation. 

For RSGC1, the spindown limit for a neutron star gravitational wave has a strain of

\begin{equation}\label{scaledspindownlimit}
    h \leq 1.54 \times 10^{-24}\left(\frac{6.6\;\rm{kpc}}{D}\right)
    \sqrt{\left(\frac{300\;\rm{yr}}{\tau}\right)\left(\frac{I_{zz}}{1.51 \times 10^{38} \rm{kg\;m^2}}\right)}
\end{equation}

benchmarked with the physical values and assumptions of this system.
Whether or not this potential signal is distinguishable from the noise of the detector determines whether a limiting statement can be made.  

To search for continuous waves, archival LIGO data can be coherently added to enhance detector sensitivity. The longer the coherence time, the deeper one may probe in frequency-amplitude space. This is similar to keeping a telescope aperture to collect more flux. This forms a 95\% confidence limit of detectability 

\begin{equation}\label{detectorstrain}
    h^{95\%}_0 = \Theta\sqrt{S_h(f)/T_{\rm{dat}}}
\end{equation}

where $S_h(f)$ is the strain noise power spectral density, $T_{\rm{dat}}$ is the data livetime, and $\Theta$ is a pipeline-dependent statistical threshold factor \citep{CasAMethods}. This statistical factor depends on the parameter space and the details of the search pipeline, and accounts for look-elsewhere effects caused by the high number of frequency/spindown pairs tested at each sky location. For a directed continuous wave search, $\Theta$ is typically in the mid-30s, and can be taken to be 35. 

In Figure \ref{fig:result}, the Advanced LIGO design curve \cite{aLIGOdesign} is integrated and scaled as if it was running for five continuous days, and the spindown limit for RSGC1 is overlaid as a horizontal line. At design sensitivity, Advanced LIGO would be capable of setting limits on continuous gravitational wave emission from RSGC1 at frequencies greater than 20 Hz, corresponding to pulsar rotations of 10 Hz and above. 

Rotating neutron star gravitational waves have the advantage of having a persistent, well-modeled signal, and integration of long stretches of data can enhance network sensitivity to rotating neutron star gravitational waves in a way that cannot be done for transient events. However, this integration is extremely computationally intensive; the standard coherent algorithm has a computing cost that scales with coherence time of $T^7$, while increasing sensitivity to gravitational wave signals as a function of $T^{1/2}$ \citep{Prix2009}.

Continuous gravitational waves are therefore just as reliant on detector improvements as transient gravitational wave events. The current Advanced LIGO configurations have not yet achieved their design sensitivity. There is currently a funded A+ upgrade to improve the sensitivity further, and designs for future interferometer networks are actively being pursued. Since known pulsars have been observed to have gravitational wave emission below the level of the spindown limit, enhancements to instrument sensitivity are necessary to facilitate future pulsar detection. We show that the current instruments can set limits on continuous gravitational wave emission from candidate T\.ZOs; improvements to instruments will extend the distance at which we can place such limits, and increase the probability of achieving a multi-messenger T\.ZO detection.

\section{Results}\label{section:results}

\begin{figure}
    \centering
    \includegraphics[width=0.48\textwidth]{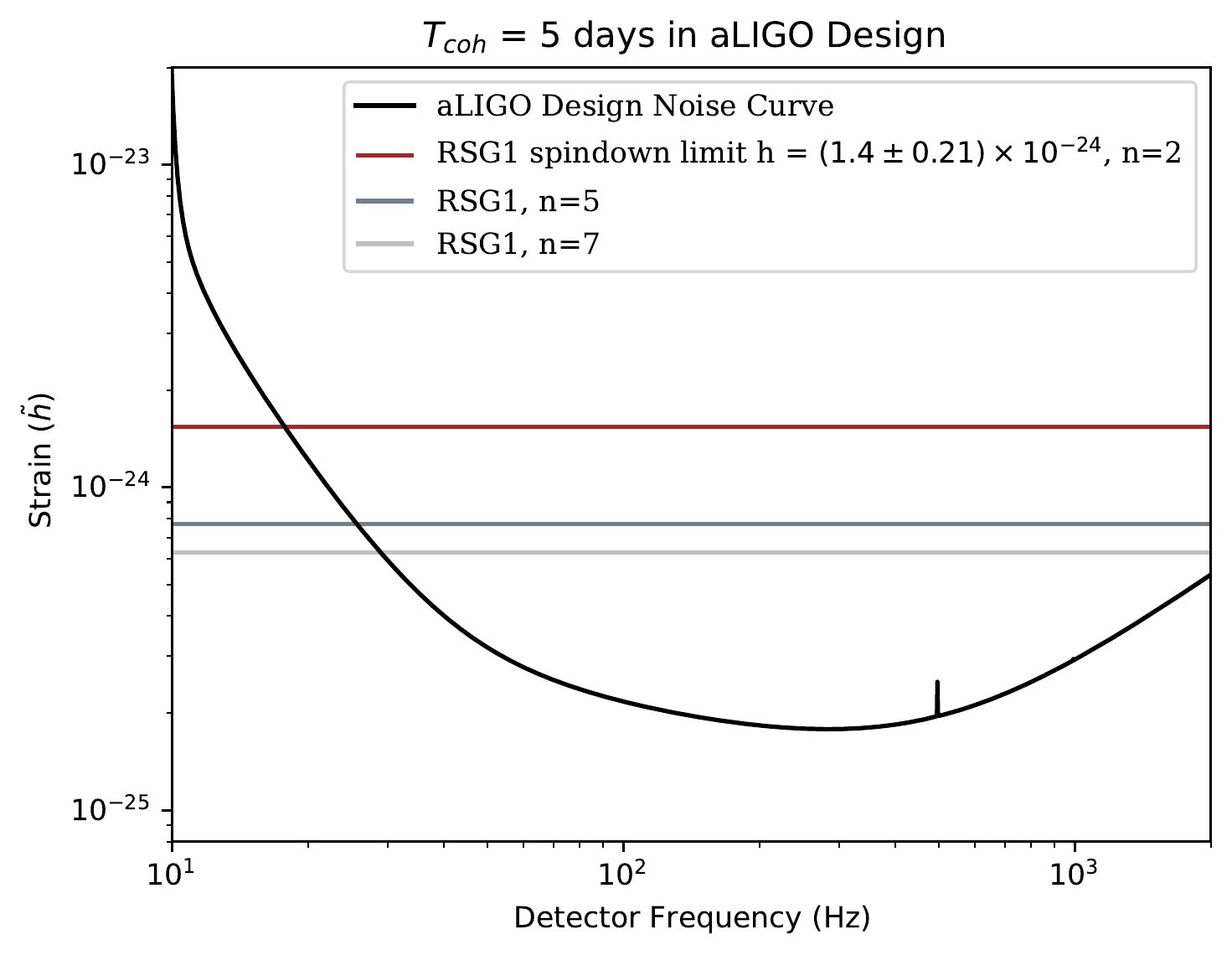}
    \caption{The aLIGO design sensitivity curve, integrated over 5 coherent days (black). The solid red line shows the strain amplitude, adopting $n=2$, of RSGC1, the most likely nearby (6.6 kpc) supergiant cluster to contain a T\.ZO observable by LIGO. For comparison, we have also included the $n=5$ (dark gray) and $n=7$ (light gray) cases.}
    \label{fig:result}
\end{figure}

The strain amplitude of LIGO detectors is subject to stochastic, constantly fluctuating noise at a given moment. Therefore, we present the strain calculated for RSGC1, a value of $1.54(\pm0.21) \times 10^{-24}$, on the characteristic noise curve of Hanford and Livingston at aLIGO design sensitivity integrated over 5 continuous days (Figure \ref{fig:result}).
This calculation follows Equation \ref{scaledspindownlimit} using the benchmarked values, representing the stiffest EoS not ruled out by GW170817 (called H4), a spindown age of 300 years, a spindown rate of $n=2$, and the distance to RSGC1 (6.6 kpc). We have scaled this equation by these values to allow the reader to easily adjust the strain amplitude for a given object.

We plot the strain of our exemplary RSG cluster as a horizontal line spanning the entire frequency range of the x-axis in Figure \ref{fig:result} in order to accommodate for different spin frequencies across a diverse spread of known pulsars \citep{Abbott_2019}. It is likely that a sufficiently braked neutron star at the center of a RSG would have a rotational frequency that places the object in the tens of Hz in the LIGO detector; it would not resemble a recycled puslar nor a millisecond pulsar. We present this work as further motivation for future detectors to improve sensitivity at low frequencies.

A strain greater than the noise floor, yet less than these upper limits is required to satisfy the GW portion of a multi-messenger T\.ZO detection. If paired with a coincident sky localization in the direction of an RSG cluster, an event would convincingly indicate a possible detection of a T\.ZO, to be further confirmed by its unique spectroscopic signature. Signals between the horizontal line in Figure ~\ref{fig:result} and the noise curve can encapsulate variations to our assumptions such as softer equations of state, less-preferred inclination angles, larger distances, and older spindown ages, but the exact configuration of these causes would be impossible to disentangle from a sole GW detection without an EM signature. It is outside the scope of this paper to perform the search itself in current released LIGO data; such a search would be tractable with application of extent methods and cluster computing.

Therefore, we point to the region in amplitude-frequency space between the noise curve and the horizontal line of RSGC1 (Figure \ref{fig:result}) for any potential CW detections of a T\.ZO (bearing in mind its sky localization and EM classification are necessary). 
Calculations of all six clusters considered in this paper were conducted, and we found all fall within this range.

\section{Conclusion}

T\.ZOs represent the truest duality of multi-messenger objects. To an electromagnetic astronomer, they appear as an RSG at first glance. Careful inspection of spectroscopic lines reveal elements that can only be fused between an envelope and the surface conditions of a neutron star core. Through the perspective of a GW detector, T\.ZOs appear similarly to pulsars- a slowly decelerating neutron star. 

In this paper, we present the most likely RSG cluster within 10kpc to harbor a T\.ZO within: RSGC1. Located 6.6kpc from Earth, this cluster has already proven to contain the ingredients necessary for T\.ZO creation- a dozen RSGs and at least 1 neutron star- making it the most well-primed candidate for observation of an already-formed T\.ZO.

Assuming a hard equation of state (H4), the typical braking index of a pulsar (n=2), and knowledge of distance to the cluster, we place upper limits on the strain needed to be observed in a ground-based GW detector to capture the continuous wave of such an object. Producing a strain of $1.54(\pm0.21) \times 10^{-24}$, a T\.ZO within RSGC1 is feasibly observable using current LIGO detectors, given a CW search with a coherence time of 5 days or greater. To allow for a full frequency range of possible neutron star spin periods, we plot a horizontal line at this strain value in Figure \ref{fig:result}. Below this line, a detected T\.ZO could represent any myriad of combinations of distances, EoS, and braking indexes that would be impossible to disentangle for certain. A coincident EM observation could help to constrain some of these parameters, and certainly to verify it was a detection of a T\.ZO at that sky location. This line may be adjusted (with the aid of Equation \ref{scaledspindownlimit}) for other candidates. It is in the scope of further work to perform this search in archival data.

\acknowledgments

The authors would like to thank D. Finstad for his invaluably fruitful and productive conversations throughout this process. EML's work on this was supported by NSF grant AST 1714285 and an award from the University of Washington Royalty Research Fund. L.D. is grateful for the financial support of the IDEAS Fellowship, a research traineeship program funded by the National Science Foundation under grant DGE-1450006.


\end{document}